\documentclass[%
a4paper,
superscriptaddress,
notitlepage,
 amsmath,amssymb,
 aps,
10pt,
twocolumn,
prl
]{revtex4-2}
 \usepackage{xcolor}
 \colorlet{mylinkcolor}{blue!66!black!80}
 \colorlet{mynew}{black}

\newcommand{\e}{{\rm e}}

\usepackage{mathtools}
 \usepackage[colorlinks=true,linkcolor=mylinkcolor,citecolor=mylinkcolor,filecolor=cyan,urlcolor=mylinkcolor,breaklinks=true]{hyperref}

 \usepackage[utf8]{inputenc}

\begin{document}
  \title{Comment on ``Inferring broken detailed balance in the absence of observable currents''}
\author{David Hartich}
\email{david.hartich@mpinat.mpg.de}
\affiliation{%
Mathematical bioPhysics Group, Max Planck Institute for
Multidisciplinary Sciences, 37077 Göttingen, Germany}
 \author{Aljaž Godec}%
 \email{agodec@mpinat.mpg.de}
\affiliation{%
Mathematical bioPhysics Group, Max Planck Institute for
Multidisciplinary Sciences, 37077 Göttingen, Germany}
\begin{abstract}
     We present a simple biophysical example that
     invalidates the main conclusion of ``Nat. Commun. 10, 3542
     (2019)''. Moreover, we explain that 
     systems with
     one or more hidden states between at least one pair of observed
     states that give rise to non-instantaneous
     transition paths
     between these states
     also invalidate the main conclusion of the aforementioned
     work. This provides a flexible roadmap for constructing 
     counterexamples. We hope for this comment to raise awareness of
     possibly hidden transition paths and of the importance
     of considering the microscopic origin of emerging non-Markovian
     (or Markovian) dynamics for thermodynamics.
\end{abstract}
\maketitle
 Ref.~\cite{mart19a}  entitled ``Inferring broken detailed balance in the absence of observable currents''  claims to derive a method
 which allows to identify  an ``underlying nonequilibrium process,
 even if no net current, flow, or drift, are present''.   
 Below we explain that the above main result of said work, which is supposed to hold for semi-Markov processes, was in fact
 never tested by the authors, nor applied to an example. 
 Remarkably, the central result of an older work by Wang and Qian \cite{wang07}
 (cited in Ref. [50] in \cite{mart19a}) already disproves the main
 conclusion of Ref.~\cite{mart19a} and, moreover, contains
a recipe to construct counterexamples to the findings of Ref.~\cite{mart19a}.

\textit{The main result of Ref.~\cite{mart19a} was never tested.---} As the main result Ref.~\cite{mart19a} derives
Eq.~(4) quantifying ``irreversibility of stationary trajectories with zero current'', which is supposed to hold for semi-Markov processes
\cite{mart19a} and is to be used to infer broken detailed balance.
While the work contains two explicit examples, none of them in fact
seems to apply to the main result. 
Instead, a variant of the main result (i.e. Eq.~(6)) is derived, which
provides a technique to identify irreversibility in certain
\emph{second order} semi-Markov processes, and is used in Figs.~2 and
~3 of \cite{mart19a}. Notably, the authors use in Eqs.~(4) and ~(6)
the same 
notation, which makes it difficult to actually notice that the
main result (that is, Eq.~(4)) was never applied to an example.

\begin{figure}
\centering
\includegraphics{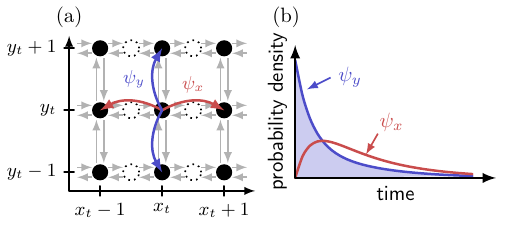}
	\caption{Motor moving in two dimensions. (a)~The filled
          circles are the positions where we can observe the motor,
          whereas it cannot be detected in the dotted states. The
          last-visited observed node forms a semi-Markov
          process. (b)~$\psi_x(t)$ and $\psi_y(t)$,
          the probability densities of the waiting time
          until the last-visited node changes in the $x$ and $y$
          direction, respectively, given in Eq.~\eqref{eq:waiting_time}.}
	\label{fig:example}
\end{figure}

\textit{Counterexample.---}Following \cite{wang07} it is straightforward to construct an example that disproves the main conclusion of \cite{mart19a}. Consider a molecular motor that walks in two directions $x$ and $y$ ``fueled'' by a chemical reaction $\text{A}\rightleftharpoons \text{B}$.
Along the $x$ direction the motor's position $x_t$  at any time $t$
advances in a two step reaction
\begin{equation}
 \mathrm{E}+\mathrm{A}+x_t\xrightleftharpoons[\kappa_1^-]{\kappa_1^+} \mathrm{EA}\xrightleftharpoons[\omega_1^-]{\omega_1^+} \mathrm{E}+ \mathrm{B} +(x_t+1),
 \label{eq:intermediate}
\end{equation}
where ``E'' represents the enzymatic motor. The intermediate step -- the
formation of the complex ``EA'' -- is assumed \emph{not} to be visible
or monitored (see dotted  circles in
Fig.~\ref{fig:example}(a)). \textcolor{mynew}{Note that one could also
  consider a more complex enzyme with more unobserved intermediate
  enzymatic states. For simplicity and without loss of generality, we stick to this simple model.}
The motor's position along the $y$ direction, $y_t$, evolves as a one-step process
\begin{equation}
 \mathrm{E}+\mathrm{A}+y_t\xrightleftharpoons[\kappa_2^-]{\kappa_2^+} \mathrm{E}+\mathrm{B}+(y_t+1).
  \label{eq:instantaneous}
\end{equation}
Whenever a molecule ``A'' is bound to the enzymatic motor we assume that the motor cannot be detected \emph{and}
the reaction \eqref{eq:instantaneous} is switched off. For simplicity
we set the rates equal to $\kappa_1^\pm=\omega_1^\pm=2$ and
$\kappa_2^\pm=1$, tacitly assuming that all chemical species are kept
at the same chemical potential, $\mu_{\rm A}=\mu_{\rm B}$. That is, the
system satisfies detailed balance \textcolor{mynew}{due to $\ln(\kappa_1^+/\kappa_1^-)=\ln(\kappa_2^+/\kappa_2^-)=\ln(\omega_1^+/\omega_1^-)=0$} and the full dynamics is a
Markov-jump process (see Fig.~\ref{fig:example}a).  The last monitored position $z_t=(x_t,y_t)$ at
any time $t$  becomes a semi-Markov process due to the hidden
intermediate step in Eq.~\eqref{eq:intermediate}.  The process $z_t$ thus
satisfies \emph{all} the assumptions that were made in
Ref.~\cite{mart19a} to derive the main
result. \textcolor{mynew}{Consequently, our example also satsfies
  Eq.~(1) in Ref.~\cite{mart19a}.} \textcolor{mynew}{Note that we refer
  to \emph{hidden states} if they are unobserved. The number of
  unobserved states in general cannot be known. In the present example
  the reader only knows the number of hidden states because we
  describe the complete underlying mathematical model for the sake of reproducibility.}

The probabilities of the two reactions to be completed for the first time are
equal, $p_{x,+}=p_{x,-}=p_{y,+}=p_{y,-}=1/4$, in both reactions.
The turnover time until any of the two reactions is completed is distributed according to waiting time densities
$\psi_{x,\pm}(t)$ and $\psi_{y,\pm}(t)$, which read
\begin{equation}
\begin{aligned}
 \psi_x(t)&=\psi_{x,\pm}(t)=\frac{4}{3}\big[2\e^{-2t}-2\e^{-8t}\big],\\
  \psi_y(t)&=\psi_{y,\pm}(t)=\frac{4}{3}\big[\e^{-2t}+2\e^{-8t}\big].
\end{aligned}
\label{eq:waiting_time}
\end{equation}
\textcolor{mynew}{To obtain  Eq.~\eqref{eq:waiting_time} 
  we solve the conditional first passage problem in
  Fig.~\ref{fig:example}, where we use a filled circle as the starting point and impose absorbing boundaries on the filled circles adjacent to the starting point (see arrows in Fig.~\ref{fig:example}).}
In this particular case the time to leave a state in either $x$ or $y$ direction
is actually exponential $\psi^{\rm
  exit}(t)\equiv\sum_{i\in \{x,y\}}[p_{i,+}+p_{i,-}]\psi_i(t)=2\e^{-2t}$ with mean
exit time $\mathcal{T}\equiv \int_0^{\infty} t \psi^{\rm exit}(t)dt=1/2$.   
As shown in Fig.~\ref{fig:example}(b) we have $\psi_x(t)\neq\psi_y(t)$, that
is, the waiting time densities  are different and thus have a genuinely non-zero
Kullback-Leibler divergence
\begin{equation}
 D_{\rm KL}[\psi_x|\psi_y]\approx 0.28,\quad D_{\rm KL}[\psi_y|\psi_x]\approx 0.47,
\end{equation}
respectively, where we have defined $D_{\rm KL}[\psi_i|\psi_j]\equiv \int_0^\infty \psi_i(t)\ln[\psi_i(t)/\psi_j(t)] dt$, where $i,j=x,y$.
Using Eqs.~(1) and (4) in \cite{mart19a} along with $D_{\rm
  KL}[\psi_i|\psi_j]\neq0$  for $i\neq j$ one would mistakenly confuse this
\emph{equilibrium system} to be out of equilibrium.  More precisely,
Eq.~(4) in \cite{mart19a}  states that the entropy production of the
waiting time is given by  $\dot S _{\rm
  WTD}=\sum_{i,j,\alpha,\beta}p_{i,\alpha}p_{j,\beta}D_{\rm
  KL}[\psi_i|\psi_j]/\mathcal{T}\approx 2\times(0.28+0.47)=1.5>0$,
where and $\alpha,\beta=\pm$; each term in the sum in fact
is non-negative. The main result of Ref.~\cite{mart19a} erroneously
predicts this equilibrium system to break detailed balance (based on
$\dot S _{\rm WTD}\neq0$).  Since our example clearly violates the
main result of Ref.~\cite{mart19a} while satisfying all the required
assumptions, we have hereby 
disproved the main result in
Ref.~\cite{mart19a}.

\textit{Opposing views on broken detailed balance.---}A similar counterexample was sketched in an earlier work by Wang and
 Qian \cite{wang07} who also stated in their abstract:
 ``We show that for a semi-Markov process detailed balance is only a
 necessary condition, but not sufficient, for its time reversibility''
 \cite{wang07}. In technical terms, Ref.~\cite{wang07} showed that if
 the waiting-time distribution to another state depends on the final
 state, the process becomes (mathematically) irreversible (here $\psi_x(t)\neq\psi_y(t)$) \emph{even if detailed balance is satisfied}. Thus, $\psi_x(t)\neq\psi_y(t)$
 must \emph{not} be used as a signature of broken detailed balance
 as erroneously concluded in Ref.~\cite{mart19a}. 
 In light of these diametrically opposing views in \cite{mart19a}
 and \cite{wang07} it is puzzling that Ref.~\cite{mart19a} actually cites
Ref.~\cite{wang07} as Reference [50].

\textit{Crucial elements of the counterexample.---}Counter\-examples to the main conclusion of \cite{mart19a} are obtained
as soon as hidden states states (see dotted circles in
Fig.~\ref{fig:example}a) 
emerge between at least one pair of observed states (see filled circles).
 This allows for passages over hidden states, called transition paths
 (see, e.g., \cite{maka21}), to become non-instantaneous.  In this
 case the coarse-graining must \emph{not} commute with the time
 reversal -- a phenomenon that we coined
 ``kinetic hysteresis'' which is an overdamped analogue of the odd
 parity of momenta \cite{hart21}.  We are not aware of any
 example without kinetic hysteresis which would allow for a
 non-vanishing waiting-time entropy production, i.e. $\dot S _{\rm
   WTD}\neq 0$, for a semi-Markov process.  For example, in absence of
 hidden cycles at least one transition-path time
 must be non-vanishing to allow for the waiting-time distribution to
 couple to the state change \cite{hart21} and in turn to allow for $\dot S _{\rm
   WTD}\neq 0$.  
Thus, if the transition paths become (effectively)
instantaneous, the waiting time does \emph{not} couple to the state
change \cite{hart21} (see \cite{erte21} for a generalization that
includes hidden cycles), i.e., $\dot S _{\rm WTD}=0$. These examples
satisfying $\dot S _{\rm WTD}=0$ clearly cannot be used to infer
``broken detailed balance in the absence of observable
currents''\textcolor{mynew}{, which was recently confirmed in Ref.~\cite{meer22} (see paragraph after Eq.~(58) therein)}.

We emphasize that while our example proves the main conclusion of
Ref.~\cite{mart19a} to be false, we do \emph{not} claim that all
mathematical results in \cite{mart19a} are incorrect.  In particular,
we explicitly acknowledge that the compact expressions for the
Kullback-Leibler divergence between two path measures indeed have some
mathematical appeal (see also Ref.~\cite{gira03}).
\textcolor{mynew}{In other words Eq.~(4) 
is  mathematically sound it does \emph{not}, however, allow to infer
broken detailed balance.}  
Moreover, whether a model in fact exists that upon
coarse-graining yields a semi-Markov process \footnote{Note
that we explicitly refer to
semi-Markov processes (sMP) and \emph{not} to possible
generalizations to
second- or higher order 
sMPs.
However, any
sMP is also a higher order sMP but the converse is untrue.}
\emph{and} concurrently allows to
infer broken detailed balance
according to \cite{mart19a} remains an intriguing question.

\textit{Conclusion.---}
\textcolor{mynew}{The authors never tested their main result Eq.~(4), which is supposed to detect broken detailed balance.}
Here we presented an explicit model of a molecular motor that
disproves \textcolor{mynew}{this} main conclusion of Ref.~\cite{mart19a}, which was in fact
already invalidated earlier in Ref.~\cite{wang07} (Ref.~[50] in
\cite{mart19a}). 
Note that the example
in Fig.~\ref{fig:example} only formally disproves the variant of the
main result in Eq.~(6) in \cite{mart19a}, which \emph{does} hold for
the more specialized
semi-Markov processes \emph{of second order} --processes with a
waiting-time distribution that depends
on the past \emph{and} future states (see
e.g. \cite{mart19a,ehri21,hart21arxiv}).  This variant thus allows to find
examples for which it holds and thus potentially allows to infer
``broken detailed balance in the absence of observable currents''
\cite{mart19a,ehri21}. 
We hope that this comment raises awareness of the importance of
considering the underlying dynamics from which possible non-Markovian
(or Markovian) coarse-grained dynamics emerge.

%
\textbf{Acknowledgments.}
 The financial support from the German Research Foundation (DFG) through the Emmy Noether Program GO 2762/1-1 to A. G. is gratefully acknowledged.
 \vspace*{-1mm}

%

 \end{document}